\documentclass[conference]{IEEEtran}
\IEEEoverridecommandlockouts
\usepackage{cite}
\usepackage{amsmath,amssymb,amsfonts}
\usepackage{algorithmic}
\usepackage{graphicx}
\usepackage{textcomp}
\usepackage{xcolor}
\def\BibTeX{{\rm B\kern-.05em{\sc i\kern-.025em b}\kern-.08em
    T\kern-.1667em\lower.7ex\hbox{E}\kern-.125emX}}

\usepackage{multirow}
\usepackage{booktabs}
\usepackage{dsfont}
\usepackage{xcolor}
\usepackage{graphicx}
\usepackage{amsmath}
\usepackage{enumerate}
\usepackage[shortlabels]{enumitem}
\usepackage{mathtools}
\usepackage{graphicx}
\usepackage{caption}
\usepackage{subcaption}
\usepackage{dblfloatfix}
\usepackage{balance}
\usepackage{soul} 

\usepackage[linesnumbered,ruled]{algorithm2e}

\begin{document}

\title{\huge Silence Speaks Volumes: Re-weighting Techniques for Under-Represented Users in Fake News Detection
}
\author{\IEEEauthorblockN{Mansooreh Karami\IEEEauthorrefmark{1}, David Mosallanezhad\IEEEauthorrefmark{2}, Paras Sheth\IEEEauthorrefmark{1}, Huan Liu\IEEEauthorrefmark{1}}
\IEEEauthorblockA{\IEEEauthorrefmark{1}Arizona State University, \IEEEauthorrefmark{2}NVIDIA}
\IEEEauthorblockA{\IEEEauthorrefmark{1}\{mkarami, psheth5, huanliu\}@asu.edu, \IEEEauthorrefmark{2}dmosallanezh@nvidia.com}
}

\maketitle

\begin{abstract}
Social media platforms provide a rich environment for analyzing user behavior. Recently, deep learning-based methods have been a mainstream approach for social media analysis models involving complex patterns. However, these methods are susceptible to biases in the training data, such as \textit{participation inequality}. Basically, a mere 1\% of users generate the majority of the content on social networking sites, while the remaining users, though engaged to varying degrees, tend to be less active in content creation and largely silent. These silent users consume and listen to information that is propagated on the platform.
However, their voice, attitude, and interests are not reflected in the online content, making the decision of the current methods predisposed towards the opinion of the active users. So models can mistake the loudest users for the majority. We propose to leverage re-weighting techniques to make the silent majority heard, and in turn, investigate whether the cues from these users can improve the performance of the current models for the downstream task of fake news detection.
\end{abstract}



\section{Introduction}
In an age where people’s opinions are often crowdsourced on Online Social Networks (OSN), a wide variety of methods have been proposed to extract patterns from these data for different tasks, such as fake news detection~\cite{karami2021profiling,cardaioli2020fake}, hate speech detection~\cite{sheth2023peace}, and recommendations~\cite{sheth2022causal,sheth2023causal}. Moreover, deep learning methods have recently become prevalent due to their ability to model the complex and non-linear relations between the input data. However, despite all the attempts to analyze social media data, these models are prone to various biases, such as \textit{participation inequality}. The participation inequality states that only a small subset of all the users usually account for a disproportionately large amount of content creation activities in social networks. 
\begin{figure}[ht]
    \centering
    \includegraphics[width=\linewidth]{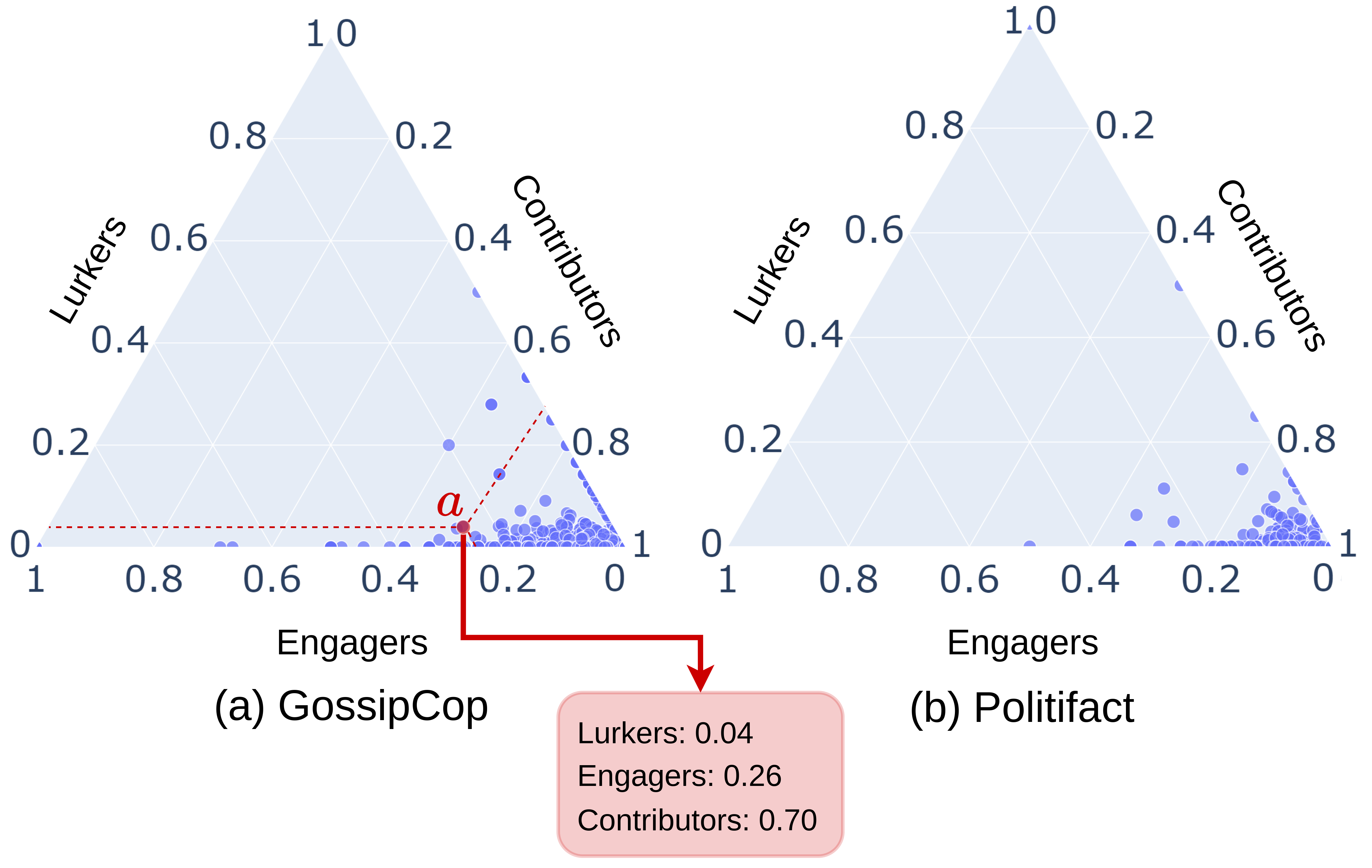}
    \caption{\small{Ternary plots of the percentage of the interactions on social media created by each of the lurker, engager, and contributor groups in fake news datasets: (a)~GossipCop and (b)~Politifact. In general, the percentage of the interactions recorded by the contributors is more than the other two groups. Out of the users who reacted to the news $a$, 4\% are lurkers, 26\% are engagers, and 70\% are contributors.}}
    \label{fig:LECcount}
\end{figure}
This phenomenon has been observed among OSN users and can be easily categorized into three types:~(1)~\textit{lurkers} who comprise 90\% of the OSN users and hardly ever participate in creating the content on social media~($\sim$1\% of the postings),~(2)~ \textit{engagers} group that contain 9\% of the social media users who occasionally contribute to content creation~($\sim$9\% of the postings), and~(3)~\textit{contributors}, who are only 1\% of the OSN users but are responsible for more than 90\% of the created content on social media. This phenomenon, which is also known as the \textit{90-9-1 Rule} or \textit{1\% Rule} by web usability experts~\cite{nielsen}, demonstrates the biases in the data that are used in current social media analysis applications.

Deep learning methods utilize the observed data to infer user behavior. However, since contributors generate most of the data, the inferred user behavior is inclined towards these users and cannot represent that of the remaining categories of users. Figure~\ref{fig:LECcount} shows the percentage of the interactions by each group of users - lurkers, engagers, and contributors - for two different datasets. 
For example, a data point (i.e., a news piece) in the lower right corner suggests that 100\% of the interactions with the news on social media are from contributors and 0\% from lurkers and engagers, respectively. 

Early studies in behavioral and social science literature often associate lurkers with names such as \textit{passive actors}~\cite{hemmings2017evaluation}, \textit{abusers of common good}~\cite{amichai2016psychological}, and \textit{free-riders}~\cite{nonnecke2003silent} that only consume resources without giving back to the community. This also influences machine learning researchers to overlook the contributions of the lurkers. However, we argue that lurkers' behavior can provide additional cues for social media analysis methods as these users actively consume and listen to the relevant information, create connections, and are receptive~\cite{edelmann2013reviewing,gong2015characterizing}. This can be corroborated by recent efforts to drive user participation in online social communities.
For example, among reasons listed in~\cite{nonnecke2001lurkers} for the lurking behavior, a user’s motivation to post is decreased if they are not able to offer any vital or novel information. Furthermore, the authors in~\cite{nguyen2022turning} mention that one of the reasons a lurker becomes active on social networking sites is when they can gain knowledge as well as propagate it outside the community. 

\begin{figure}[ht]
    \centering
    \includegraphics[width=0.85\linewidth]{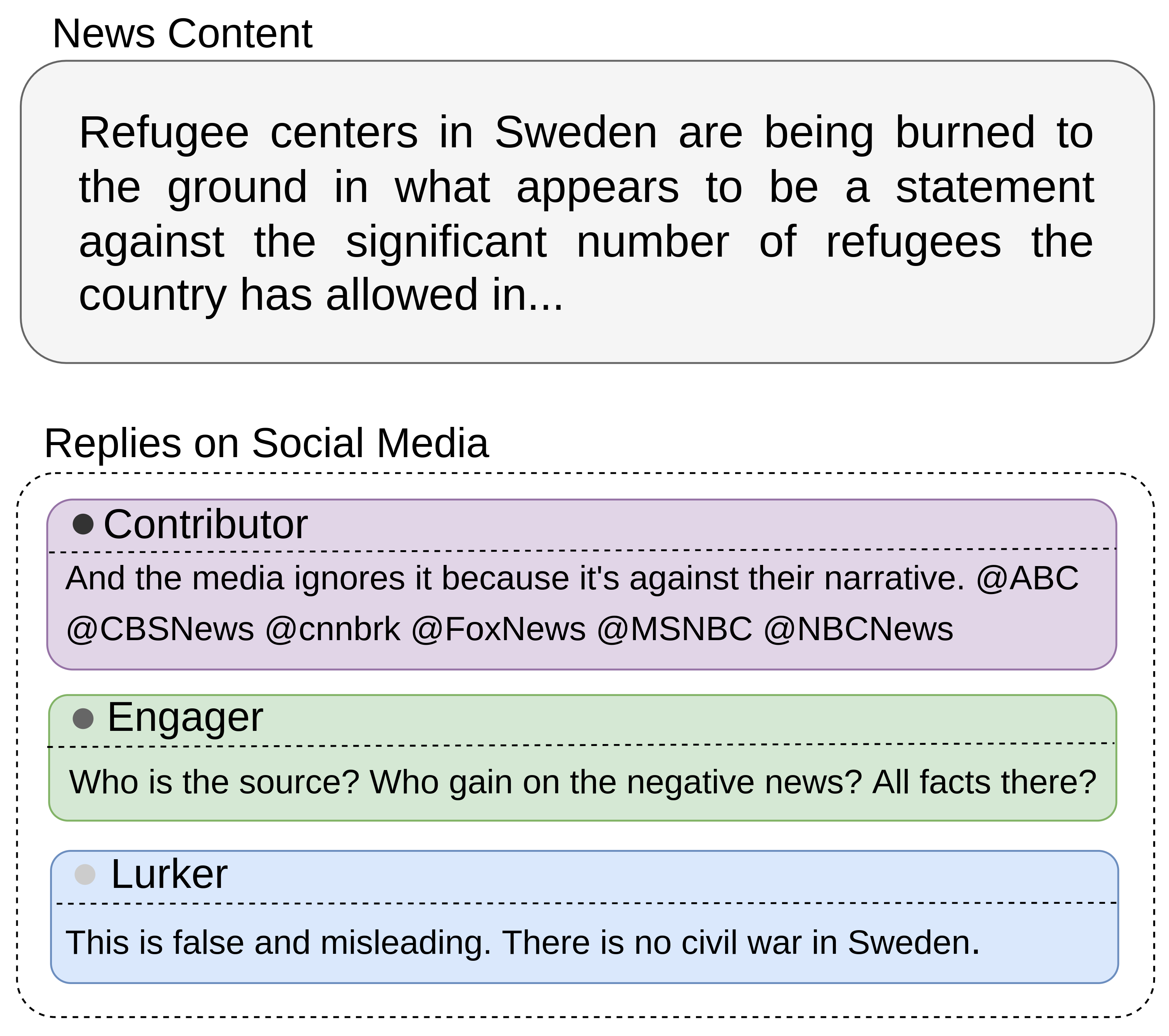}
    \caption{\small{Example of a piece of news content from the Politifact dataset showing the tweets of users from each group
    . We hypothesize that if a piece of news provokes a lurker to create content on social media, giving importance to such interaction might improve the performance of fake news detection models.}}
    \label{fig:hypothesis}
\end{figure}

Given these reasons, a lurker might engage with a post when they have valuable information to add related to the topic. Thus, we hypothesize that giving importance to such interactions between the posts and lurkers may improve the performance of the different social media analysis applications.
For instance, consider the task of fake news detection. This task entails classifying a news article as real or fake by benefiting from the user-news interactions obtained from social media data. However, directly utilizing this network may not be fruitful due to two reasons. First, as mentioned, this interaction may be biased toward the views of the contributors as they are the ones creating about 90\% of the interactions. Second, unobserved interactions (i.e., unshared news) do not guarantee that the user was not exposed to the news. A user might be exposed to the article but may choose to refrain from expressing their opinions due to one or more reasons. For example, a user might doubt the post's veracity or a user may feel like they might not add value to the already propagated content. In compliance with the earlier stated hypothesis, if a lurker engages with a news article, they might have more information about the news article. Thus, by up-weighting the limited lurkers' interaction, one may improve the detection capabilities of the fake news detection model. 
Figure~\ref{fig:hypothesis} shows a motivational example from the Politifact dataset that includes fact-checked news articles. The example includes the content of the fake news and different tweets that mention the news from three different types of users. In this example, the news provoked the lurker to comment on its falsity.
In this work, we only utilized retweet interactions.

We propose to leverage re-weighting techniques to verify whether silence speaks volumes. We use the task of fake news detection and evaluate its performance by differentiating between various interactions based on the user categories. Our approach learns a representation that reflects the actual landscape of the platform and assigns higher weights to news that triggered the silent users more, as they could potentially offer additional information for fake news detection.



The main contribution of this work is three-fold:
\begin{itemize}
	\item To the best of our knowledge, this is the first attempt to consider the types of users based on their activity for fake news detection.
	\item We design, implement, and experiment with two weighting techniques to upvalue the under-represented users on social media platforms and record the performance for the downstream task.
	\item We extend two benchmark datasets in the field of fake news detection to also include the information of a user being a lurker, engager, and contributor which can be utilized for generalized user behavioral analysis.
\end{itemize}

\section{Related Work}
The proposed methodology spans the subject domains of online participation, class imbalance, and fake news detection. The state-of-the-art in these areas is discussed in this section.
\subsection{Online Participation and Lurking Behavior}
In the field of psychology, behavioral, and social science, there is a wide range of studies dedicated to extracting factors that drive user participation as well as lurking behavior in online social communities. These behavioral factors can be classified into three major categories: (1)~individual-level,
(2)~community-level,
and (3)~environmental-level.
Note that we did not include offline barriers such as \textit{user's available time} since they were not directly associated with lurking behavior.

\subsubsection{Individual-level Factors} Studies suggest that demographic features such as gender and age as well as personality traits play an important role in online participation~\cite{liu2017big, nonnecke2001lurkers}. Four prevailing intrinsic characteristics are (1)~\textit{extraversion} that captures quantity and intensity of interpersonal interactions, (2)~\textit{neuroticism} that captures susceptibility to emotional instability, (3)~\textit{narcissism} that captures excessive self-promotional behavior, and (4)~\textit{self-efficacy} that captures self-confidence in one’s own ability to successfully accomplish specific tasks or achieve desired outcomes.

\subsubsection{Community-level Factors}

The prominent factor related to social and community for online participation is the \textit{social identity}. Social identity is defined as how people perceive themselves as a part of a particular community~\cite{mousavi2017interpreting}. In other words, members share information to obtain a sense of belonging and identification.
Another influence is the \textit{reciprocity} factor that looks into how much the community can provide for its members as well as how much an individual can return the benefits and reduce the perceived indebtedness~\cite{sun2014understanding, nguyen2022turning}.

\subsubsection{Environmental-level Factors}
The most influential factor in the active participation of social media users related to the platform is the high \textit{perceived ease of use} which is defined as the degree to which the technology is easily understood~\cite{nonnecke2006non}.
On the other hand, the ease of use should not result in limited functionality of the platform as it would lead to a decrease in user engagement. Other factors include the privacy-preserving functionality and security-related issues of the platform.

There are also multiple factors that would involve two or more of the above categories such as \textit{privacy} and \textit{security} of the communities as well as the platform.
Nevertheless, if lurkers decide to break their voices, the above factors might play an important role. Out of which the need of giving back to the community~(i.e., \textit{reciprocity}) and the confidence in possessing the knowledge to contribute to the online content~(i.e., \textit{self-efficacy}) are the core motivations of this paper. 

\subsection{Class Imbalance and Long-Tail Distribution}
\label{subs:classImb}

The natural data classes exhibit a long-tail distribution in which the sample counts across classes are imbalanced. In other words, there are a few classes with a large number of samples while most of the other classes include a relatively fewer number of examples. This poses a challenge as most models are typically trained on artificially balanced datasets, making them vulnerable in practice when applied to real-world data. Various approaches have been developed to address this performance bias, which can be broadly categorized into three groups: (1)~re-sampling approaches that involve either under-sampling the majority class or over-sampling the minority class~\cite{chawla2004special}, (2)~re-weighting methods that apply cost-sensitive learning or loss re-weighting for different classes or different samples~\cite{cui2019class}, and (3)~augmentation-based methods in which they artificially expand the dataset by applying transformation functions on the data samples~\cite{bhattacharjee2022text}.




The concept of imbalancedness in this work is similar to class imbalance problems but varies in terms of its source. In this paper, the task classes are different from the users' participation inequality. For example, sentences extracted from social media for the task of sentiment analysis can be balanced (or imbalanced) in terms of the number of samples for positive, negative, and neutral classes; while the number of users for each user type based on the intensity of their activity who created these sentences still be highly skewed.

\subsection{Disinformation Spreader and Fake News Detection}
In the field of user-based fake news detection and fake news spreader profiling, researchers have utilized different conjunctions of user's profile information, user's activity, user's network connectivity, and user's generated content~\cite{antelmi2019characterizing,karami2021profiling,cheng2021causal}. Cheng et al.~\cite{cheng2021causal} proposed a model to identify the causal relationships between users' profiles and their susceptibility to sharing fake news articles. The authors modeled the dissemination of fake news by creating implicit feedback based on the user's exposure and interest in specific fake news. The learned fake news sharing behavior is then used in improving the detection of fake news. Karami et al.~\cite{karami2021profiling} extracted some features from the user's profile information, generated content, and activity that represents their motivational behavior in spreading fake news. They showed the effectiveness of their model in determining which users are more likely to spread fake news. Cardaioli et al.~\cite{cardaioli2020fake} investigated how the behavioral-based features such as Big Five personality and stylometric features extracted from the content of a user's timeline can be used to profile fake news spreaders. Shu et al.~\cite{shu2019role} investigated the importance of explicit features such as register time, follower and following count as well as implicit user meta information such as location and political bias inferred from their online behaviors and historical tweets for the detection. 

Nevertheless, all the aforementioned methods do not distinguish between lurkers, engagers, and contributors, hence, generalizing the dissemination behavior for all types of users.

\begin{figure*}[ht]
    \centering
    \includegraphics[width=0.85\linewidth]{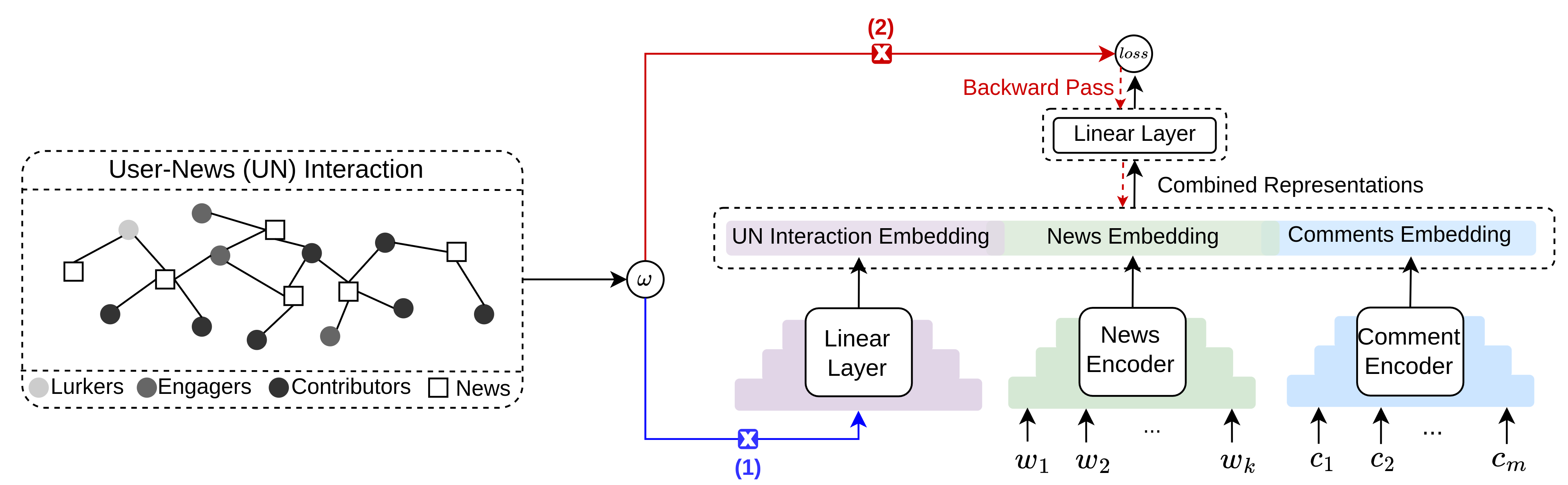}
    \caption{\small{Two re-weighting strategies were used to learn a balanced representation for the task of fake news detection: {\color{blue}(1)~Edge Re-weighting ($\mathcal{x}$\ref{subsubsec:edge})} and {\color{red}(2)~Sample-level Re-weighting ($\mathcal{x}$\ref{subsubsec:sample})}.}}
    \label{fig:arch}
\end{figure*}

\section{Problem Statement}
Let $\mathcal{X}=\{(x_1,y_1), (x_2, y_2), ..., (x_n, y_n)\}$ denote a set of $n$ news articles with labels $y=0$ for true and $y=1$ for fake news. Each news article $x_i$ consists of three components: (1)~the news content, $a_i \in \mathcal{A}$, which is a sequence of $k$ words $\{w_1, w_2, ..., w_k\}$, (2)~a set of $m$ comments containing different views of the users' opinion related to the corresponding news article, $c_i = \{c_{1i}, c_{2i}, ..., c_{mi}\} \in \mathcal{C}$, and (3)~a user-news interaction $u_{ji}\in \mathcal{U}$ with $p$ number of users.

Typically, $\mathcal{U}$ is a \textit{binary} matrix representing interaction between user $j$ and news $i$: if $j$ interacts with $i$ then $u_{ji} = 1$, otherwise $u_{ji} = 0$. Note that $u_{ji} = 0$ can be interpreted as either the user $j$ was not exposed to the news article $i$ or was exposed to but due to some reasons (e.g., not sure of the veracity of the news~\cite{cui2021voice}) chose not to propagate it. Based on our hypothesis, to investigate the impact of interactions with under-represented users, we aim to design a fake news detection function that considers the type of users in terms of their activity, $\mathcal{G}=\{L, E, C\}$.

Formally, we can represent the model as follows:

\begin{center}
\fbox{\parbox[c]{0.95\linewidth}{Given news articles $\mathcal{A}$, users' comments $\mathcal{C}$, and a user-news interaction $\mathcal{U}$, learn a fake news detection function $f(\mathcal{A}, \mathcal{C}, \mathcal{U}, \mathcal{G})\rightarrow \hat{y}$ with respect to the users belonging to one of the lurkers~(L), engagers~(E), and contributors~(C) groups $\mathcal{G}$.}}
\end{center}

\section{Designing Fake News Detection Model}
\label{sec::fnd}
Previous methods in fake news detection either do not consider user-news interaction in their model, or it is appended as a binary matrix with 1 showing the user tweeted or retweeted about specific news. Similar to other social media analysis studies, this news dissemination data in online environments is also biased toward the users who create the majority of the social media content. In other words, the user-news interaction matrix is biased towards the views of the users that are more eager on asserting their opinion about the news but belong to only 1\% of the social media population - i.e. the contributors. The focus of this paper is to provide a fair representation by giving more value to the interactions created by lurkers.

We design two approaches (Figure~\ref{fig:arch}). The first method balances the user-news interaction matrix which later will be added to the baseline models as a weighted matrix. The second method will apply sample re-weighting based on the activity of the users to see whether this would improve the performance of the downstream task. 

In this section, we will briefly talk about the text representation learning for news articles as well as the news comments and then introduce our weighting mechanisms.

\subsection{News Articles and Users' Comments Representations}
To generate a vector representation of the news content as well as the users' comments, different models apply different text representations. In the task of fake news detection, earlier methods use word-level and sentence-level features such as bag-of-words and n-grams. Recent models use deep learning-based methods such as Recurrent neural networks (RNN), Long Short Term Memory (LSTM), and Transformers to model sequential data. Transformers use a self-attention mechanism to extract vital information from the input data. Both the news and the comment encoder inputs are text sequences, and they output the vector representation of text. Formally, if we show the article's content and the comment encoder as $g_a(\cdot)$ and $g_c(\cdot)$ functions, respectively, then for each news $i$,
\begin{equation}
\label{eq:encoder}
\begin{multlined}
    z_{ia} = g_a(w_1, w_2, ..., w_k) \quad \text{and} \quad z_{ic} = g_c(c_{1}, c_{3}, ..., c_{m})
\end{multlined}
\end{equation}
where $z_{ia}$ and $z_{ic}$ are the embedding vectors for the news content and the user comments, respectively, $w_1, w_2, ..., w_k$ is the sequence of the words in the news articles and $c_{1}, c_{2}, ..., c_{m}$ are its corresponding comments.

\subsection{Edge Re-weighting Mechanism for News Dissemination Network}
\label{subsubsec:edge}
\begin{figure}[htb]
    \centering
    \includegraphics[width=\linewidth]{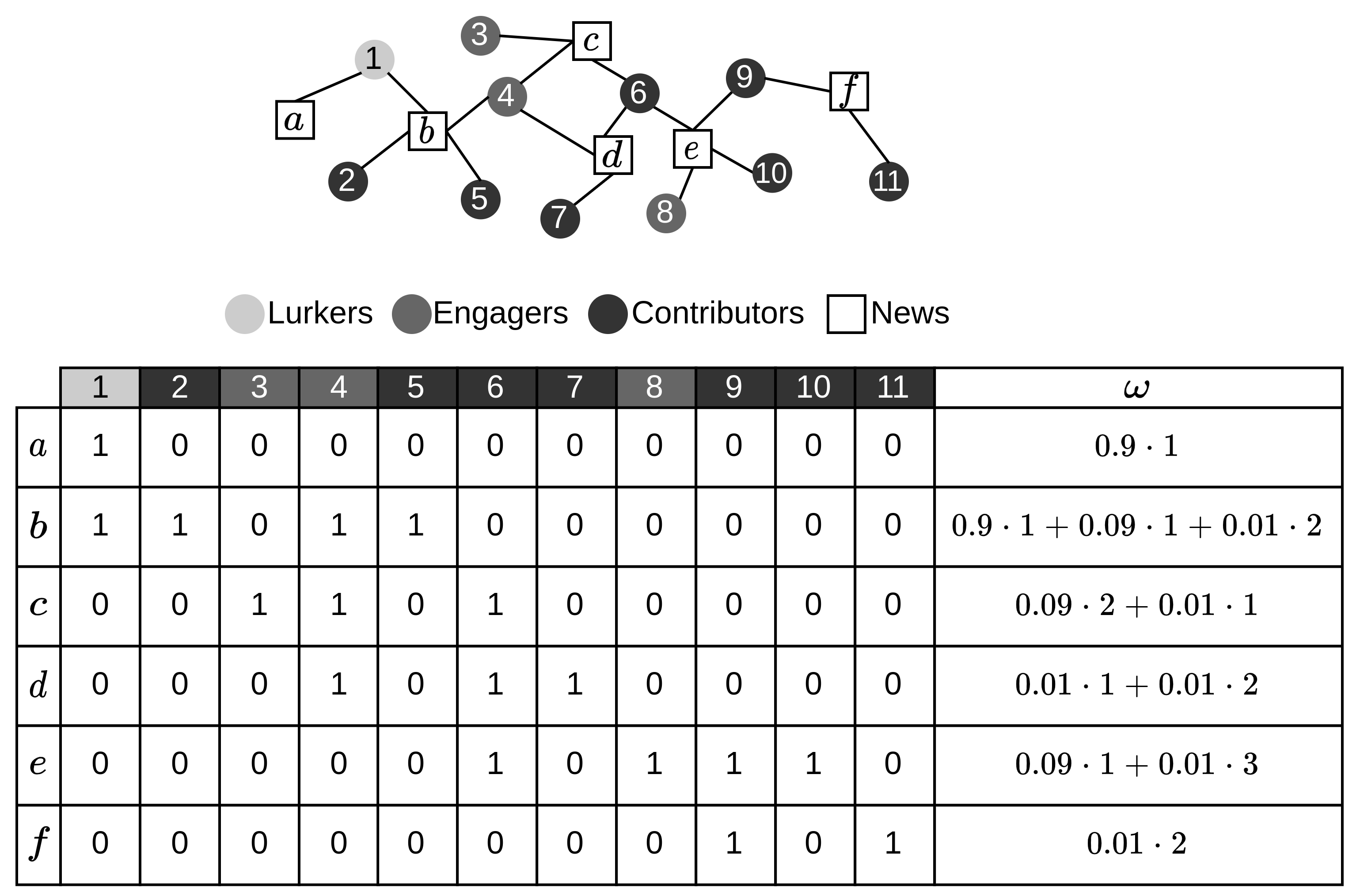}
    \caption{\small{An example of a network with 11 users (1 lurker, 3 engagers, and 7 contributors) interacting with 6 pieces of news. This interaction vector is a binary vector with 1 indicating the existence of an interaction. The weights are calculated based on equation~\ref{eq:reweight2}.}}
    
    \label{fig:weightexample}
\end{figure}

The news dissemination network consists of two different types of nodes: users and news. In Figure~\ref{fig:weightexample}, users are denoted by circles while the news pieces are illustrated by squares. Each user node can belong to one category of lurkers, engagers, or contributors.

To handle the imbalancedness of the user types on social media, we propose a weighting mechanism based on the 90-9-1 Rule. The calculated weight would be applied to all the edges connected to a square-shaped node based on the type of all its connected circle-shaped nodes. Formally, we substitute the binary user-news interaction matrix~$(\mathcal{U})$ in our formulation of the fake news detection function with a normalized weighted version~$(\mathcal{\overline{U}})$. We propose the following weighting mechanism:
\begin{equation}
\label{eq:reweight}
    \overline{u}_i = u_i \cdot  \left(1+\frac{\omega_i}{\parallel\omega\parallel}\right)^{\alpha}  \quad {\forall i \in\{1,...,n\}}
\end{equation}
where $\omega_i$ is calculated as follows:
\begin{equation}
\begin{aligned}
    \label{eq:reweight2}
     \omega_i & = [0.9\cdot\sum_{j=1}^p \mathds{1}_L (j) \cdot u_{ji}
      + 0.09\cdot\sum_{j=1}^p \mathds{1}_E (j) \cdot u_{ji}\\
      & + 0.01\cdot\sum_{j=1}^p \mathds{1}_C (j) \cdot u_{ji}]
\end{aligned}
\end{equation}
 In the above equations, $u_i$ is a vector showing the user's interaction activity (i.e., 0 or 1) with all the news. $n$ and $p$ are the number of news articles and users, respectively. $L$, $E$, and $C$ are the list of lurkers, engagers, and contributors. The $\alpha\geq 0$ is a hyperparameter that controls the intensity of the weighting mechanism. For example, $\alpha = 1$ will apply a weighting based on the 90-9-1 Rule on each user type while $\alpha = \frac{1}{2}$ is the smoother version of it. Moreover, $\mathds{1}_S (j)$ is an indicator function and is 1 if $j \in S$, otherwise, it is 0, where $S$ is one of the user types. The indicator functions defines which type a specific user belongs to. An example is given in Figure~\ref{fig:weightexample}. In this figure, for instance, four users interacted with news $b$, out of which one is a lurker, one is an engager, and two are contributors. The weight is calculated as:
 \begin{equation}
 \begin{aligned}
     \omega_b & = 0.9\cdot\text{(\# of lurkers)} + 0.09\cdot\text{(\# of engagers)} \\ 
     & + 0.01\cdot\text{(\# of contributors)} = 0.9\cdot 1 + 0.09 \cdot 1 \\
     &+ 0.01 \cdot 2 = 1.01
 \end{aligned}
 \end{equation}

\subsection{Sample-level Re-weighting Mechanism for News Representation}
\label{subsubsec:sample}

Sample re-weighting has been a mainstream approach in creating a robust model when dealing with imbalanced training data~\cite{cao2019learning,cui2019class}. Inspired by this, we trained the models by applying a sample-level re-weighting method based on the users belonging to lurker, engager, or contributor groups. In other words, for the news article $i$ and $M$ number of samples in a batch, the normalized weight is integrated into the loss function to model a balanced fake news detection. Formally,
\begin{equation}
    \label{eq:balanced_loss}
    \mathcal{L}_{balanced} = -\frac{1}{M} \sum_{i=1}^{M} \left(1+\frac{\omega_i}{\parallel\omega\parallel}\right)^\alpha \cdot \mathcal{L}_{CE}(y_i, \hat{y}_i)
\end{equation}
where $y_i$ and $\hat{y}_i$ is the true and the predicted labels, respectively.
The weights are calculated as a batch-wise version of equation~\ref{eq:reweight2}. Moreover, $\mathcal{L}_{CE}$ is the cross-entropy loss, formulated as:
\begin{equation}
    \label{eq:ce_loss}
    \mathcal{L}_{CE}(y_i, \hat{y}_i) = y_i \log (\hat{y}_i) + (1-y_i) \log (1-\hat{y}_i)
\end{equation}
\noindent
The batch-wise learning process of the balanced fake news detection and the weighting procedure is provided in Algorithm~\ref{alg:slrw}.

\section{Experimental Setting}
In this section, we describe the details of the experimental setup including the benchmark datasets, dataset preparation, baseline methods, and implementation details.
\subsection{Datasets and Dataset Preparation}
We used two datasets from the FakeNewsNet repository as the seed datasets for the evaluation: Politifact and GossipCop~\cite{shu2020fakenewsnet}.
\begin{itemize}
    \item Politifact\footnote{https://www.politifact.com/}: a fact-checking website where reporters and editors from the media fact-check political news articles. The URLs of news articles are available on the Politifact website and are used to collect tweets related to them.
    \item GossipCop\footnote{https://www.gossipcop.com/}: a website for fact-checking entertainment stories aggregated from various media outlets. On the GossipCop website, articles get a score between 0 and 10 as the degree from fake to real.
\end{itemize}

\begin{algorithm}[ht]
\small
	\caption{Learning to Re-weight News Representations Based on the User Types.}\label{alg:slrw}
	\SetKwInOut{Input}{Input}
    \SetKwInOut{Output}{Output} 
    
    \Input{$\mathcal{X}_{tr}$; $\theta^0$; \text{epochs}; UN Matrix $[u_{ji}]$; $\alpha$; Lurkers~(L), Engagers~(E), and Contributors~(C) sets.}
    \Output{$\theta^T$}
    \textbf{for} $e=0, ..., \text{epochs}$ \textbf{do}:\\
    $\qquad$\textbf{for} $t=0, ..., T-1$ \textbf{do}:\\
    $\qquad\qquad\mathcal{X}^t_{tr}\leftarrow$ SampleMiniBatch$(\mathcal{X}_{tr}, t)$\\
    $\qquad\qquad \hat{y}^t_{tr}\leftarrow$ Forward$(\mathcal{X}^t_{tr},\theta^t)$\\
    $\qquad\qquad \omega^t_{tr}\leftarrow \sum_{S\in\{L, E, C\}}w_S\cdot\sum_{j=1}^p \mathds{1}_S (j) \cdot u_{ji}
    $\\
    $\qquad\qquad loss = mean\left[\left( 1+\frac{\omega^t_{tr}}{\parallel\omega\parallel}\right)^\alpha\mathcal{L}_{CE}(y_{tr}^t, \hat{y}_{tr}^t)\right]$\\
    $\qquad\qquad \nabla\theta^t\leftarrow$ Backward$(loss, \theta^t)$\\
    $\qquad\qquad \theta^{t+1}\leftarrow$ OptimizerStep$(\theta^t, \nabla\theta^t)$\\
    $\qquad$\textbf{end for}\\
    \textbf{end for}
\end{algorithm}

In these datasets, along with the content of the news, the news comments and IDs of the Twitter users who reposted these fake and real stories are also included. The textual data (i.e., news content and news comments) were pre-processed to remove punctuation, out-of-vocabulary words, URLs, hashtags, and mentions. We utilized the Twitter user ids to create the user-news interaction matrix.

We also collected the history of the activities of each of the Twitter users identified in the Politifact and GossipCop datasets. Some of these users were deleted or suspended accounts and we were not able to access their activity and profile information anymore (9,537 of the GossipCop users and 13,181 of the Politifact users). We ignored these users in our matrix creation. For the rest, to categorize them into three groups of lurkers, engagers, and contributors, we calculate the average number of activities per day. We set the thresholds for the average number of activities per day in creating the lurkers and engagers to 0.025 and 0.15, respectively, such that it approximately follows the 90-9-1 Rule~\cite{nielsen} as well as the definition provided in social science behavioral papers~\cite{sun2014understanding}. Statistics of the created datasets are summarized in Table~\ref{tab:stat}.

\begin{table}[htb]
\centering
\small
\caption{Statistics of the Datasets.}
\label{tab:stat}
\begin{tabular}{clcc} 
\toprule
\multicolumn{2}{l}{} & \multicolumn{1}{l}{\textbf{Politifact}} & \textbf{GossipCop} \\ 
\hline\hline
\multirow{3}{*}{\begin{tabular}[c]{@{}c@{}}\textbf{Number of}\\\textbf{News}\end{tabular}} & \textbf{\textit{Real}} & 132 & 3,588 \\
 & \textbf{\textit{Fake}} & 319 &  2,230 \\ 
 \cline{2-4}
& \textbf{Total} & 451 &  5,818 \\ 
\hline\hline
\multirow{4}{*}{\begin{tabular}[c]{@{}c@{}}\textbf{Number of}\\\textbf{Interactions}\end{tabular}} & \textbf{\textit{Lurkers}} & 482 & 382 \\
 & \textbf{\textit{Engagers}} & 
 4,295 & 3,945  \\
 & \textbf{\textit{Contributors}} & 
  41,738 & 30,054 \\ 
  \cline{2-4}
  & \textbf{Total} & 46,515 & 34,381 \\ 
\hline\hline
\textbf{\# of Comments} &  & 89,999 & 231,269  \\
\bottomrule
\end{tabular}
\end{table}

\subsection{Baselines}
\label{subsec:basel}
In this section, for evaluation, we consider state-of-the-art baselines that use both news content and users' comments. To also include the \textsc{BERT}~\cite{kenton2019bert} model to the group of baselines, we integrate \textsc{BERT} with a comment encoder for a fair comparison. 

The followings are the details regarding each baseline:
\begin{itemize}
    \item \textsc{CSI}~\cite{ruchansky2017csi}: This method applies a hybrid deep model to capture the characteristics of fake news such as the text of the article, the set of tweets in which users commented about the fake news, and the source of the article such as the credibility of the media source. For a fair comparison, we disregarded the news source feature.

    \item d\textsc{EFEND}~\cite{shu2019defend}: This model applies deep hierarchical sentence-comment co-attention network. d\textsc{EFEND} learns feature representations of the content and the comments for fake news detection and jointly discovers explainable sentences from these two sources.
    
    \item \textsc{TCNN-URG}~\cite{qian2018neural}: Based on convolutional neural network idea for text classification~\cite{kim2014convolutional}, this model tries to capture semantic information from the article's text using Two-level Convolutional Neural Network (\textsc{TCNN}). Moreover, it incorporates a User Response Generator (\textsc{URG}) module to learn a variational autoencoder to model the user responses to the article and generate responses for unseen news articles.
    
    \item \textsc{BERT+HAN}: We created a variant of the \textsc{BERT} model that includes the comments to match the other baseline models. We added the Hierarchical Attention Network for training the news comment section following Mosallanezhad et al.~\cite{mosallanezhad2022domain} which models the importance of each comment along with the salient word features.
\end{itemize}

\begin{table*}
\centering
\small
\caption{The performance on the original architecture of the baselines along with a variation that includes the binary user-news interaction component (+UN) as well as variations that incorporate the proposed re-weighting techniques (i.e., user-news edge re-weighting and sample re-weighting methods). The highest accuracy is bolded for each row.}
\label{table:performance}
\begin{tabular}{ccccccccc} 
\toprule
\multirow{2}{*}{\textbf{Dataset}} & \multirow{2}{*}{\textbf{Model}} & Original &  & \multicolumn{1}{c}{\begin{tabular}[c]{@{}c@{}}With Binary User-News \\Interaction Module (+UN)\end{tabular}} &  & Edge Re-weighting &  & Sample Re-weighting \\ 
\cline{3-3}\cline{5-5}\cline{7-7}\cline{9-9}
 &  & \textbf{Accuracy} &  & \textbf{Accuracy} &  & \textbf{Accuracy} &  & \textbf{Accuracy} \\ 
\hline \hline
\multirow{4}{*}{\textbf{Politifact}} & CSI & 81.10 $\pm$ 1.07 &  & 85.93 $\pm$ 2.63 &  & \textbf{87.25 $\pm$ 1.40} &  & 86.59 $\pm$ 1.76 \\ 
\cline{2-9}
 & dEFEND & 81.48 $\pm$ 1.50 &  & 84.36 $\pm$ 2.20 &  & 86.72 $\pm$ 0.72 &  & \textbf{87.16 $\pm$ 1.51}   \\ 
\cline{2-9}
 & TCNN-URG & 80.32 $\pm$ 2.06 &  & 86.92 $\pm$ 1.24 &  & \textbf{92.41 $\pm$ 2.22} &  & 88.57 $\pm$ 0.53 \\ 
\cline{2-9}
 & BERT+HAN & 83.04 $\pm$ 1.35 &  & 87.25 $\pm$ 1.32 &   & \textbf{89.67 $\pm$ 0.80} &   &  88.79 $\pm$ 0.82\\ 
\hline\hline
\multirow{4}{*}{\textbf{GossipCop}} & CSI & 85.98 $\pm$ 0.29 &  & 88.77 $\pm$ 0.50 &  & \textbf{91.13 $\pm$ 0.42} &  & 89.94 $\pm$ 0.74 \\ 
\cline{2-9}
 & dEFEND & 78.34 $\pm$ 1.55 &  & 87.62 $\pm$ 0.84 &  & \textbf{88.81 $\pm$ 0.32} &  &  88.79 $\pm$ 0.22\\ 
\cline{2-9}
 & TCNN-URG & 81.42 $\pm$ 2.62 &  & 85.66 $\pm$ 0.46 &  & \textbf{85.95 $\pm$ 0.68} &  & 85.21 $\pm$ 1.83 \\ 
\cline{2-9}
 & BERT+HAN & 71.86 $\pm$ 0.00 &  & 88.14 $\pm$ 0.41  &  & \textbf{89.21 $\pm$ 0.17} &  & 88.42 $\pm$ 0.33 \\
\bottomrule
\end{tabular}

\end{table*}

\subsection{Implementation Details}
Traditional fake news detection methods only utilize the text of the news for detecting the fake from the real. However, integrating auxiliary information would provide a comprehensive representation of the samples and help in improving the performance of the models. For example, news comments provide useful signals for fake news detection~\cite{shu2019defend, mosallanezhad2022domain}, since semantic cues such as signals supporting or doubting the veracity of the content can be extracted from the comments. On the other hand, user-news interactions can highlight the type of items a user interacts with and further improve the understanding of user behaviors~\cite{mosallanezhad2022domain, shu2019beyond, shu2022cross}. Moreover, it has been well documented that, fake news tends to spread faster than true news articles on social media sites such as twitter. Thus, incorporating user-item interactions provides additional cues to enhance fake news detection.
To study the effectiveness of our weighting mechanism in the task of fake news detection, we integrated this user-news interaction component into each of the baseline models. In other words, the output of the news and comment encoders were concatenated to the user-news interaction encoder which is a feed-forward network, and was fed to a dense layer to be trained for the fake news detection task, similar to the illustration provided in the Figure~\ref{fig:arch}. 
Table~\ref{table:performance} shows the performance (accuracy) of these models with the original architecture, when the binary user-news interaction is added, and when we incorporate the two proposed weighting techniques.

To improve the training process time of the \textsc{BERT+HAN} models, we initialize the news and comments encoder by fine-tuning them with the news content and users' comments, respectively. Due to \textsc{BERT}'s input size limitation, we truncate each news content and comment to include its first 512 words. The embedding dimension for the HAN architecture is set to 100. Both the news content and user comments networks were trained using a simple feed-forward fake news classifier on top of it which was removed in the final architecture of the model. Once pre-trained, we merged the news and comments encoders in the \textsc{BERT+HAN} model with the user-news interaction encoder. With passing the news elements (i.e., news content, user comments, and user-news interaction matrix) through this integrated network, we train the final fake news classifier.

We trained the models with early stopping for all the baselines. For the edge re-weighting mechanism, instead of the binary user-news interaction matrix, we fed the weighted version, while for the sample-level re-weighting, we changed the loss based on the equation~\ref{eq:balanced_loss}. Moreover, we tracked all the experiments using the Weights \& Biases tool~\cite{wandb} where applicable. The hyperparameters tuned are the batch size, epochs, and learning rate.

\section{Experimental Results}
In this section, we review the designed experiments using the task of fake news detection. We specifically are looking to answer the following research questions:
\begin{enumerate}[leftmargin=.5in, label={\bfseries Q\arabic*.}]
        \item How much effect do the designed weighting mechanisms have on the performance of the models?
        \item Which weighting mechanism would capture the voice of the silence better?
\end{enumerate}
Using the available data, one way to investigate whether the voices of the silent users make a difference is to up-weight the silent users' signals and compare the performance of the downstream task with the original case. To be able to apply the weighting procedures based on the designed architecture, at first, we need to integrate the user-news interaction module (i.e., the UN interaction Embedding in Figure~\ref{fig:arch}) to different baselines introduced in $\mathcal{x}$\ref{subsec:basel} and record their performance. Comparing the first two accuracy columns in Table~\ref{table:performance}, we can see that user-news interaction conveys valuable information when added to the current fake news detection algorithms. The average improvement in the accuracy of the models for Politifact news is +4.63\% while the average improvement of +8.14\% has been observed in the GossipCop dataset.

In the following sub-sections, we investigate each of the above questions (i.e., \textbf{Q1} in $\mathcal{x}$\ref{subsec:edge-reweighting} and \textbf{Q2} in $\mathcal{x}$\ref{subsec:sample-reweighting}) along with the discussions on the results.



\subsection{How much effect do the designed weighting mechanisms have on the performance of the models?}
\label{subsec:edge-reweighting}
To check whether in fact the cues from the silent users have additional information and can improve the performance of the current models,
we will apply the proposed re-weighting techniques and look into the performance of the downstream task. With that, as our first attempt at incorporating the type of users who retweeted the news for fake news detection, we started by re-weighting the edges of the user-news network as described in section~\ref{subsubsec:edge}.
As another re-weighting technique, we added the sample-level re-weighting technique to the loss of the deep neural network to learn a re-weighting of the inputs as introduced in section~\ref{subsubsec:sample}. This technique, based on the gradient direction, learns to up-weight those news articles that provoke silent users more since they may contain additional cues for detection.

By comparing the performance values with the models with the binary user-news interaction, we can infer how much of the increase in performance is due to the weighting procedure. In other words, it will give more importance to the voice of the under-represented groups and see whether this would change the performance of the downstream task.
Overall, for all models in the edge re-weighting technique, we can see an average of +2.82\% and +1.23\% improvement for the Politifact and GossipCop datasets, respectively, when compared to the model with binary user-news interaction.  Same with the sample re-weighting technique, in which the average of +1.66\% and +0.55\% improvement has been achieved.

\begin{figure}[htb]
\centering
\begin{subfigure}{0.45\textwidth}
\centering
  \includegraphics[width=0.9\linewidth]{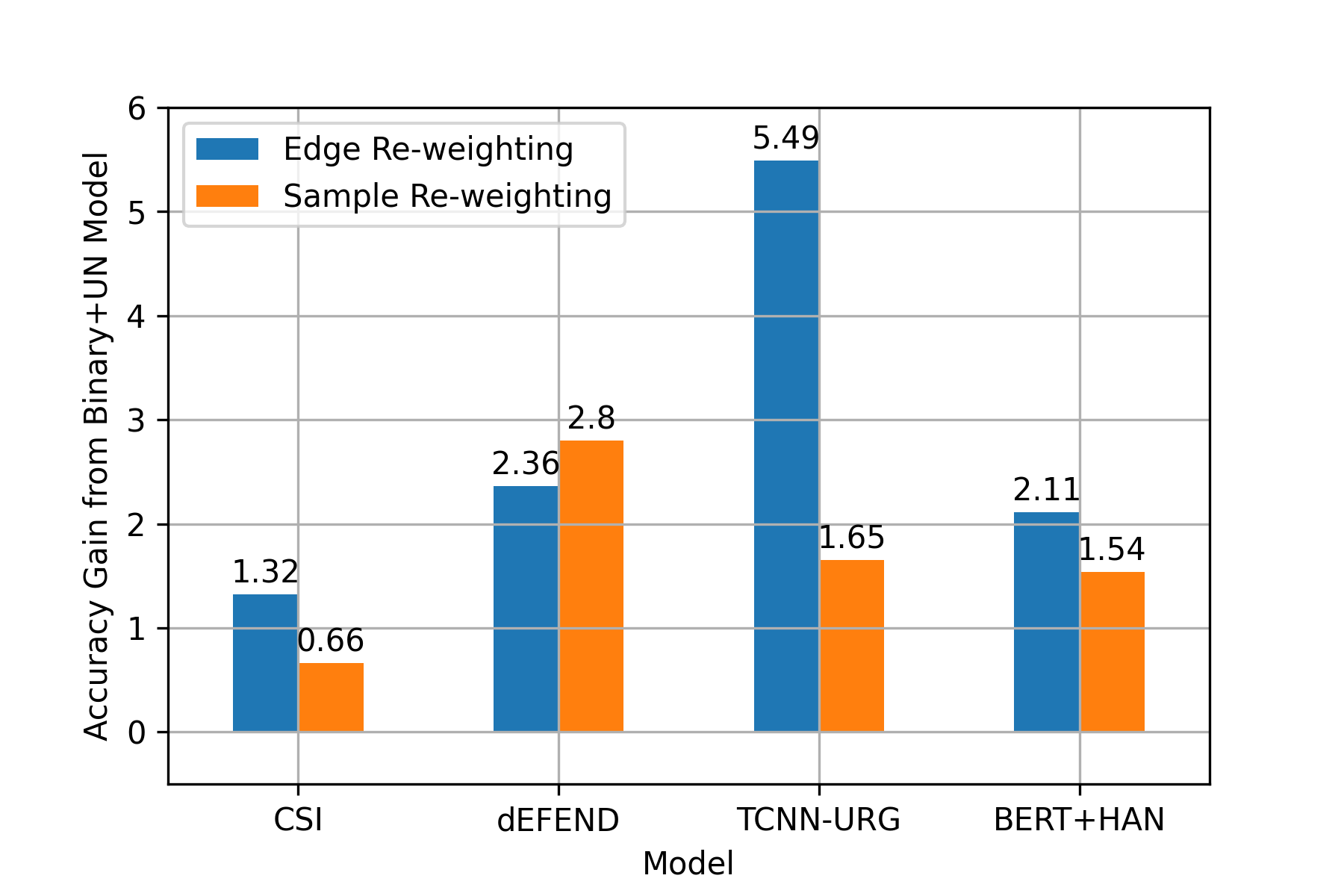}
  \caption{Politifact Dataset}
  \label{fig:sub1}
\end{subfigure}%

\begin{subfigure}{0.45\textwidth}
  \centering
  \includegraphics[width=0.9\linewidth]{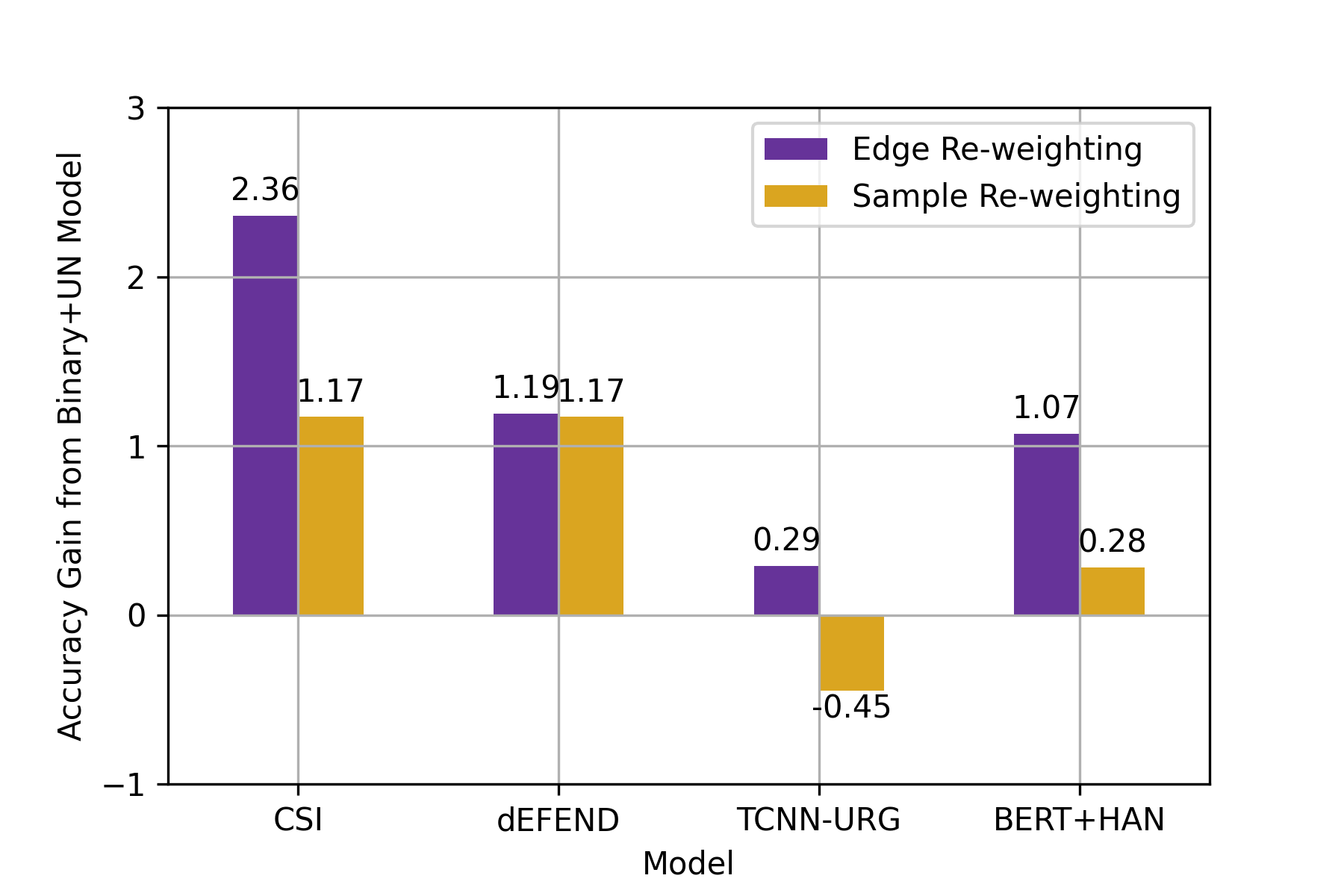}
  \caption{GossipCop Dataset}
  \label{fig:sub2}
\end{subfigure}
\caption{Accuracy gain of the proposed techniques in comparison with the model with binary UN interaction for (a)~PolitiFact and (b)~GossipCop. The edge re-weighting method has consistently yielded improvements across all the baselines.}
\label{fig:gain}
\end{figure}

In conclusion, when the results of the two techniques are compared with the original architecture of the models and with the case when the binary user-news interaction matrix is added, both techniques provide evidence to support our hypothesis. 
The improvement, although slight, can provide us with a representation that gives importance to the potential cues in silent users' interactions. 
The reason for this marginal improvement is mostly because of the limited positive interaction of the lurkers with the news. For example, out of the 34,381 users who reposted the news in the GossipCop dataset, only 382 are lurkers. Re-weighting these signals would help, but it is not expected to provide us with a significant improvement. In addition to these signals, if we were able to provide other cues such as whether a user is interested in a piece of news or topic, we would have expected to see more improvement. However, with the API limitations, such data is not accessible.

\subsection{Which weighting mechanism would capture the voice of the silence better?}
\label{subsec:sample-reweighting}
To see which weighting mechanism is better at capturing the voice of the silence, we can look into the amount of improvement with both of the models and compare them with each other. By comparing the values in each line of the Table~\ref{table:performance}, except for one case (i.e., sample re-weighting for dEFEND model in Politifact dataset), the highest accuracy has been captured by the edge-reweighting technique. To better visualize the difference, Figure~\ref{fig:gain} shows the accuracy gain for both edge-reweighting and sample re-weighting methods. By comparing both methods, on average edge re-weighting improvements were higher and more consistent among all models when compared with the sample re-weighting values. 

 As another observation, by comparing the results of the different datasets used in our experiment, the models' improvement is more evident when the number of news is limited. Despite the power of deep neural networks for text classification, their effectiveness and performance highly depend on the quantity and quality of the labeled data. As listed in Table~\ref{tab:stat}, the number of news in Politifact is 451, while the number of news in GossipCop is about 13 times more, with 5,818 pieces of news. However, the edge re-weighting technique applied to the models provided a more robust representation in the case when the number of training data is limited and scarce.

\section{Conclusion and Future Work}
In this paper, we suggest two weighting techniques to upvalue the under-represented users on social media. From our observations on the empirical results, the results of the edge re-weighting method were consistent for all the baselines and improved the accuracy of the detection.
It is worth mentioning that the assigned weights in the weighting formula can be leveled based on the platform. Since some works reported the 3-level Nielsen’s rule being extreme~\cite{antelmi2019characterizing}, with some statistical analysis, weight alignment can be applied based on the user's behavior on different platforms. Moreover, since, to the best of our knowledge, this is the first attempt in considering user types in terms of the activities, more potential solutions can be investigated. Our priority with this work is to raise the issue of \textit{participation inequality} with the currently deployed models.

In this work, due to API limitations, we only considered those users as lurkers if their minimal activity was recorded. In other words, we only examined the positive interactions and ignored negative ones (i.e., zeros in the UN matrix). Since some of the lurkers are highly active on social media (i.e., daily logins and consuming content) but do not post any content at all, future work, can exchange the user-news interaction matrix with the user's exposure matrix~\cite{karami2022estimating} and interpret the degree of interestingness of a piece of news for a user. Therefore, creating a less sparse user-news interaction matrix.

\section*{Acknowledgment}
This material is based upon work supported by ONR (N00014-21-1-4002). Opinions, interpretations, conclusions, and recommendations are those of the authors.

\balance
\bibliographystyle{ieeetr}
\bibliography{main}

\begin{thebibliography}{10}

\bibitem{karami2021profiling}
M.~Karami, T.~H. Nazer, and H.~Liu, ``Profiling fake news spreaders on social
  media through psychological and motivational factors,'' in {\em The 32nd ACM
  Conference on Hypertext and Social Media}, 2021.

\bibitem{cardaioli2020fake}
M.~Cardaioli, S.~Cecconello, M.~Conti, L.~Pajola, and F.~Turrin, ``Fake news
  spreaders profiling through behavioural analysis,'' in {\em CLEF}, 2020.

\bibitem{sheth2023peace}
P.~Sheth, T.~Kumarage, R.~Moraffah, A.~Chadha, and H.~Liu, ``Peace:
  Cross-platform hate speech detection-a causality-guided framework,'' {\em
  arXiv preprint arXiv:2306.08804}, 2023.

\bibitem{sheth2022causal}
P.~Sheth, R.~Guo, K.~Ding, L.~Cheng, K.~S. Candan, and H.~Liu, ``Causal
  disentanglement with network information for debiased recommendations,'' in
  {\em SISAP}, Springer, 2022.

\bibitem{sheth2023causal}
P.~Sheth, R.~Guo, L.~Cheng, H.~Liu, and K.~S. Candan, ``Causal disentanglement
  for implicit recommendations with network information,'' {\em ACM
  Transactions on Knowledge Discovery from Data}, 2023.

\bibitem{nielsen}
J.~Nielsen, ``Participation inequality: The 90-9-1 rule for social features,''
  2006.
\newblock Nielsen Norman Group, Accessed on 21 Sept 2021.

\bibitem{hemmings2017evaluation}
K.~Hemmings-Jarrett, J.~Jarrett, and M.~B. Blake, ``Evaluation of user
  engagement on social media to leverage active and passive communication,'' in
  {\em 2017 IEEE ICCC}, 2017.

\bibitem{amichai2016psychological}
Y.~A-Hamburger, T.~Gazit, J.~Bar-Ilan, O.~Perez, N.~Aharony, J.~Bronstein, and
  T.~Dyne, ``Psychological factors behind the lack of participation in online
  discussions,'' {\em Computers in Human Behavior}, 2016.

\bibitem{nonnecke2003silent}
B.~Nonnecke and J.~Preece, ``Silent participants: Getting to know lurkers
  better,'' in {\em From usenet to CoWebs}, pp.~110--132, Springer, 2003.

\bibitem{edelmann2013reviewing}
N.~Edelmann, ``Reviewing the definitions of “lurkers” and some implications
  for online research,'' {\em Cyberpsychology, Behavior, and Social
  Networking}, vol.~16, no.~9, pp.~645--649, 2013.

\bibitem{gong2015characterizing}
W.~Gong, E.-P. Lim, and F.~Zhu, ``Characterizing silent users in social media
  communities,'' in {\em Proceedings of the ICWSM}, vol.~9, 2015.

\bibitem{nonnecke2001lurkers}
B.~Nonnecke and J.~Preece, ``Why lurkers lurk,'' {\em Americas Conference on
  Information Systems}, 2001.

\bibitem{nguyen2022turning}
T.-M. Nguyen, L.~V. Ngo, and W.~Paramita, ``Turning lurkers into innovation
  agents: An interactionist perspective of self-determinant theory,'' {\em
  Journal of Business Research}, vol.~141, pp.~822--835, 2022.

\bibitem{liu2017big}
D.~Liu and W.~K. Campbell, ``The big five personality traits, big two
  metatraits and social media: A meta-analysis,'' {\em Journal of Research in
  Personality}, vol.~70, pp.~229--240, 2017.

\bibitem{mousavi2017interpreting}
S.~Mousavi, S.~Roper, and K.~A. Keeling, ``Interpreting social identity in
  online brand communities: Considering posters and lurkers,'' {\em Psychology
  \& Marketing}, vol.~34, no.~4, pp.~376--393, 2017.

\bibitem{sun2014understanding}
N.~Sun, P.~P.-L. Rau, and L.~Ma, ``Understanding lurkers in online communities:
  A literature review,'' {\em Computers in Human Behavior}, 2014.

\bibitem{nonnecke2006non}
B.~Nonnecke, D.~Andrews, and J.~Preece, ``Non-public and public online
  community participation: Needs, attitudes and behavior,'' {\em Electronic
  Commerce Research}, vol.~6, no.~1, pp.~7--20, 2006.

\bibitem{chawla2004special}
N.~V. Chawla, N.~Japkowicz, and A.~Kotcz, ``Special issue on learning from
  imbalanced data sets,'' {\em ACM SIGKDD}, 2004.

\bibitem{cui2019class}
Y.~Cui, M.~Jia, T.-Y. Lin, Y.~Song, and S.~Belongie, ``Class-balanced loss
  based on effective number of samples,'' in {\em Proceedings of the CVPR},
  2019.

\bibitem{bhattacharjee2022text}
A.~Bhattacharjee, M.~Karami, and H.~Liu, ``Text transformations in contrastive
  self-supervised learning: A review,'' in {\em IJCAI}, 2022.

\bibitem{antelmi2019characterizing}
A.~Antelmi, D.~Malandrino, and V.~Scarano, ``Characterizing the behavioral
  evolution of twitter users and the truth behind the 90-9-1 rule,'' in {\em
  Companion Proceedings of The WWW Conference}, 2019.

\bibitem{cheng2021causal}
L.~Cheng, R.~Guo, K.~Shu, and H.~Liu, ``Causal understanding of fake news
  dissemination on social media,'' in {\em 27th ACM SIGKDD}, 2021.

\bibitem{shu2019role}
K.~Shu, X.~Zhou, S.~Wang, R.~Zafarani, and H.~Liu, ``The role of user profiles
  for fake news detection,'' in {\em IEEE/ACM ASONAM}, 2019.

\bibitem{cui2021voice}
H.~Cui and T.~Abdelzaher, ``The voice of silence: interpreting silence in truth
  discovery on social media,'' in {\em IEEE/ACM ASONAM}, 2021.

\bibitem{cao2019learning}
K.~Cao, C.~Wei, A.~Gaidon, N.~Arechiga, and T.~Ma, ``Learning imbalanced
  datasets with label-distribution-aware margin loss,'' {\em NeurIPS}, 2019.

\bibitem{shu2020fakenewsnet}
K.~Shu, D.~Mahudeswaran, S.~Wang, D.~Lee, and H.~Liu, ``Fakenewsnet: A data
  repository with news content, social context, and spatiotemporal information
  for studying fake news on social media,'' {\em Big data}, 2020.

\bibitem{kenton2019bert}
J.~D. M.-W.~C. Kenton and L.~K. Toutanova, ``Bert: Pre-training of deep
  bidirectional transformers for language understanding,'' in {\em Proceedings
  of NAACL-HLT}, pp.~4171--4186, 2019.

\bibitem{ruchansky2017csi}
N.~Ruchansky, S.~Seo, and Y.~Liu, ``Csi: A hybrid deep model for fake news
  detection,'' in {\em Proceedings of the 2017 ACM on CIKM}, 2017.

\bibitem{shu2019defend}
K.~Shu, L.~Cui, S.~Wang, D.~Lee, and H.~Liu, ``defend: Explainable fake news
  detection,'' in {\em Proceedings of the 25th ACM SIGKDD}, 2019.

\bibitem{qian2018neural}
F.~Qian, C.~Gong, K.~Sharma, and Y.~Liu, ``Neural user response generator: Fake
  news detection with collective user intelligence.,'' in {\em IJCAI}, vol.~18,
  pp.~3834--3840, 2018.

\bibitem{kim2014convolutional}
Y.~Kim, ``Convolutional neural networks for sentence classification,'' in {\em
  Proceedings of the 2014 Conference on EMNLP}, 2014.

\bibitem{mosallanezhad2022domain}
A.~Mosallanezhad, M.~Karami, K.~Shu, M.~V. Mancenido, and H.~Liu, ``Domain
  adaptive fake news detection via reinforcement learning,'' in {\em
  Proceedings of the ACM Web Conference 2022}, pp.~3632--3640, 2022.

\bibitem{shu2019beyond}
K.~Shu, S.~Wang, and H.~Liu, ``Beyond news contents: The role of social context
  for fake news detection,'' in {\em WSDM'19}, 2019.

\bibitem{shu2022cross}
K.~Shu, A.~Mosallanezhad, and H.~Liu, ``Cross-domain fake news detection on
  social media: A context-aware adversarial approach,'' in {\em Frontiers in
  Fake Media Generation and Detection}, pp.~215--232, Springer, 2022.

\bibitem{wandb}
L.~Biewald, ``Experiment tracking with weights and biases,'' 2023.
\newblock Software available from wandb.com.

\bibitem{karami2022estimating}
M.~Karami, A.~Mosallanezhad, P.~Sheth, and H.~Liu, ``Estimating topic exposure
  for under-represented users on social media,'' {\em arXiv:2208.03796}, 2022.

\end{thebibliography}

\end{document}